\documentclass[12pt,preprint]{aastex}

\newcommand{\sect}[1]{\S\,\ref{#1}}
\newcommand{\be}{\begin{displaymath}}
\newcommand{\ee}{\end{displaymath}}
\newcommand{\bea}{\begin{eqnarray}}
\newcommand{\eea}{\end{eqnarray}}
\newcommand\msol{M_{\odot}}
\newcommand\cc{$^{12}$C/$^{13}$C}
 
\shortauthors{Denissenkov \& Pinsonneault}
\shorttitle{$^3$He-Driven Mixing In Low-Mass Red Giants}

\begin{document}

\title{$^3$HE-DRIVEN MIXING IN LOW-MASS RED GIANTS:
       CONVECTIVE INSTABILITY IN RADIATIVE AND ADIABATIC LIMITS}

\author{Pavel A. Denissenkov\altaffilmark{1,2}, and Marc Pinsonneault\altaffilmark{1}}
\altaffiltext{1}{Department of Astronomy, The Ohio State University, 4055 McPherson Laboratory,
       140 West 18th Avenue, Columbus, OH 43210; dpa@astronomy.ohio-state.edu, pinsono@astronomy.ohio-state.edu.}
\altaffiltext{2}{On leave from Sobolev Astronomical Institute of St. Petersburg State University,
   Universitetsky Pr. 28, Petrodvorets, 198504 St. Petersburg, Russia.}
 
\begin{abstract}
We examine the stability and observational consequences of
mixing induced by $^3$He burning in the envelopes of first ascent red giants.
We demonstrate that there are two unstable
modes: a rapid, nearly adiabatic mode that we cannot identify with an underlying
physical mechanism, and
a slow, nearly radiative mode that can be identified with thermohaline convection.
We present observational constraints that make the operation
of the rapid mode unlikely to occur in real stars.  Thermohaline convection turns out to be fast enough
only if fluid elements have finger-like structures with a length to diameter
ratio $l/d\ga 10$. We identify some potentially serious obstacles for thermohaline
convection as the predominant mixing mechanism for giants.
We show that rotation-induced horizontal turbulent diffusion
may suppress the $^3$He-driven thermohaline convection. Another potentially serious
problem for it is to explain observational evidence of enhanced extra mixing. 
The $^3$He exhaustion in stars approaching the red giant branch (RGB) tip
should make the $^3$He mixing inefficient on the asymptotic giant branch (AGB).
In spite of this, there are observational data indicating the presence of extra mixing
in low-mass AGB stars similar to that operating on the RGB.
Overmixing may also occur in carbon-enhanced metal-poor stars.
\end{abstract} 

 
\section{Introduction}
\label{sec:intro}

There is strong observational evidence for deep mixing in the radiative envelopes of 
low-mass ($M\la 2\,\msol$) red giant branch (hereafter, LM-RGB) stars.  
Changes in light element abundances (such as Li, C, N) and in the \cc\ ratio as 
a function of luminosity have been seen in low- and solar-metallicity red giants both 
in the field and in stellar clusters (e.g., \citealt{gb91,grea00,bea01,kea01,gea02,sh03,sm03,sb06,spiteea06}).  
The observed pattern requires at least a component of {\it in situ} mixing. 
This extra mixing could have consequences for other species (such as $^3$He) that are not directly observed.
Indeed, in spite of the predicted efficient production of $^3$He in low-mass main sequence (MS) stars, 
its Galactic abundance has been nearly constant since the epoch of Big Bang nucleosynthesis (e.g., \citealt{t98,bea02,vfea03}).
To explain this, semi-empirical stellar evolution models have shown that the carbon depletion due 
to extra mixing in LM-RGB stars should unavoidably be accompanied by 
a strong $^3$He destruction that counterbalances its production on the MS (\citealt{rea84,h95,ch95,wea96}). 

However, the true physical process that is responsible for mixing has resisted identification.  
Rotationally induced mixing has been an implied underlying mechanism since the pioneering work of 
\cite{sm79}, but there are serious difficulties in reconciling 
the observed mixing pattern with theoretical predictions (\citealt{chea05,pea06}). 
Moreover, it can be shown that rotational mixing solely dependent on the local angular velocity gradient 
$q\equiv (\partial\ln\Omega/\partial\ln r)$ (e.g., shear mixing) will be self-quenching.
Indeed, the empirically constrained mixing rate
$v_{\rm mix}\ga 10^{-3}$\,--\,$10^{-2}$\,cm\,s$^{-1}$ (see \sect{sec:concl}) is faster (as it should be)
than the mass inflow rate $|\dot{r}|\leq 10^{-4}$\,cm\,s$^{-1}$ in the radiative zones of LM-RGB stars. Hence, if $v_{\rm mix}$ were
proportional to $|q|^n\ (n>0)$ then the flattening of the rotation profile
by the accompanying angular momentum redistribution would quench the mixing very quickly.

A very different class of solution has recently been proposed by 
\cite{eea06} (hereafter, EDL06; see also \citealt{eea07}).
While investigating the core He flash in a low-mass model star near the RGB tip with the 
code {\tt Djehuty}\footnote{This is a 3D explicit hydrodynamics code with the time step constrained
by the Courant condition which can couple to a 1D stellar evolution code.}, 
\cite{dea06} noticed some gas motions in the radiative zone above the H-burning shell,
in addition to convective motions driven by $^4$He burning in the core. In their follow-up papers, 
they have made a conclusion that these additional gas motions
are due to the Rayleigh-Taylor instability (RTI) driven by $^3$He burning. They have noted
that, unlike other nuclear reactions in stars, the reaction $^3$He\,($^3$He,\,2p)$^4$He
decreases the mean molecular weight $\mu$. EDL06 have argued
that, even though the decrease of $\mu$ is minute ($|\Delta\mu|\sim 10^{-4}$), it is the resulting local inversion of
the $\mu$-gradient $\nabla_\mu \equiv (d\ln\mu/d\ln\,P)<0$ that has led to
the  RTI in their 3D simulations. They claim that velocities of
gas motions induced by the RTI are ``comparable to the velocity of the normal
convection'' ($v_{\rm c}\sim 10^5$\,cm\,s$^{-1}$) and that this is consistent with the simple heuristic estimate of
$v^2\sim gH_P(\Delta\mu/\mu)$, where $g$ and $H_P$ are the local gravity and pressure scale height.
 
\begin{figure}
\plotone{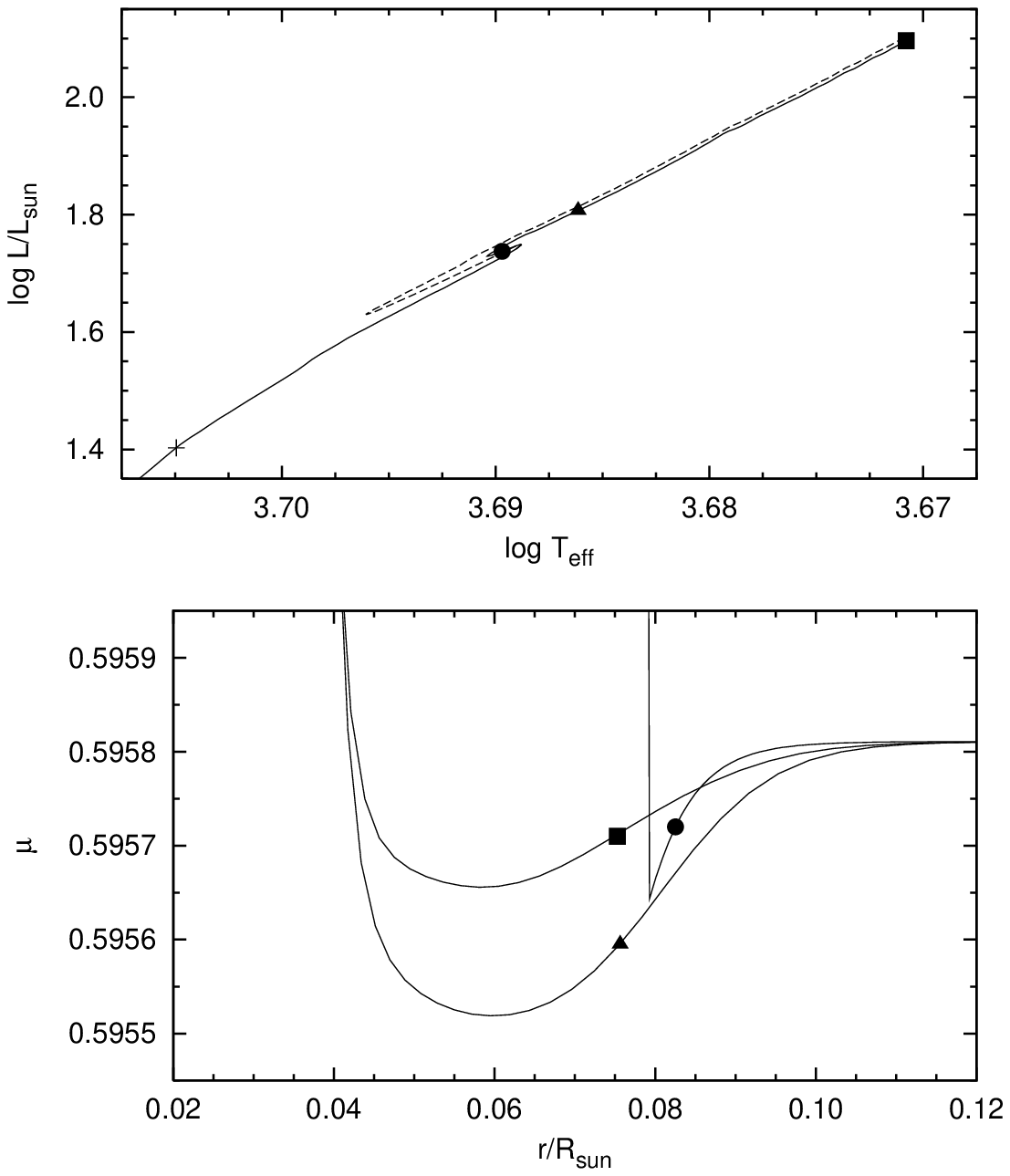}
\caption{Upper panel: evolutionary tracks of a $0.83\,\msol$ model star
         (the initial H and He mass fractions are $X=0.758$ and $Y=0.24$)
         near the bump luminosity (solid curve --- for $\Gamma = 0.01$;
         dashed curve --- for $\Gamma = 0.4$). Cross marks the end of
         the first dredge-up. Bottom panel: profiles of the mean molecular
         weight in the radiative zones of our unmixed red giant models locations of which are
         shown with the same symbols in the upper panel. Depressions of $\mu$
         are caused by the reaction $^3$He\,($^3$He,\,2p)$^4$He.
         } 
\label{fig:f1}
\end{figure}

A $\mu$-gradient inversion from $^3$He burning is a predicted consequence of standard stellar evolution.
After a low-mass star has left the MS its convective envelope first deepens (the first dredge-up) and then
its bottom starts to recede. At the depth of its maximum inward penetration the convective envelope imprints
a discontinuity in the chemical composition profile. Later on, the advancing in mass H-burning shell
will erase this discontinuity. During this event the evolution of the star slows down, which produces
bumps in differential luminosity functions of stellar clusters, and the star itself makes a tiny
zigzag on the HRD diagram (Fig.~\ref{fig:f1}, upper panel). However, before the major H-burning shell
erases the composition discontinuity, a shell in which $^3$He burns down and which advances
in front of the H-burning shell will cross the discontinuity first. The $^3$He($^3$He,2p)$^4$He reaction
reduces the mean molecular weight locally. However, this reduction is so minute, $\Delta\mu\approx\mu^2\Delta X_3/6\approx -10^{-4}$
(here, $\mu\approx 0.6$, and $X_3$ is the $^3$He mass fraction that can reach up to a value of $\sim$\,$2\times 10^{-3}$
in the envelope of a low-mass star ascending the RGB), that it is not seen on the $\mu$-profile
until the $\mu$ depression will find itself in the chemically homogeneous region of the radiative zone previously occupied
by the convective envelope.  This happens close to the bump luminosity (Fig.~\ref{fig:f1}). 

In this paper we investigate the $^3$He instability in more detail.
First of all, we note that the RTI would not be expected
in compressible, and hence stratified, stellar material.
However, a $\mu$-gradient inversion may trigger instabilities related to convection.
In section 2, we discuss general criteria for convective instability taking into consideration that,
over longer timescales, fluid elements can exchange heat with their surroundings, thus modifying the background temperature stratification.
We identify two families of solutions: a rapid mode with a nearly adiabatic
thermal structure and a slow mode with a nearly radiative structure. The EDL06 results appear
to correspond to the rapid mode,
while the slow mode can be identified as thermohaline convection (e.g., \citealt{v04}). On the basis of
previously published diffusion coefficient estimates it is likely to
be triggered in the envelopes of red giants.  However, there are significant (and
uncertain) assumptions related to the actual operation of this instability.  We discuss 
key features determining the efficiency of the slow mode: the assumed geometry of the fluid elements,
which directly impacts the timescale for exchanging heat; the potential impact of horizontal turbulence
in suppressing the instability; and the predicted depth of mixing.
In section 3 we demonstrate that the alternative approach based on the linear stability analysis of
the underlying conservation equations also leads to the conclusion that thermohaline convection
may be suppressed by the horizontal turbulent diffusion.

In section 4 we evaluate the impact of any proposed mixing mechanism on the thermal
structure of the red giant branch stars.  We argue that the rapid mode, which is similar in its properties to the originally
published results of EDL06, can be ruled out because it would induce strong feedback on the thermal
structure of giants and would predict a mixing pattern contrary to observations.  We also demonstrate
that both the slow mode and prior empirical mixing estimates would not disturb the thermal structure of
giants. In section 4 we assess the overall promise of $^3$He mixing as a mechanism. We find that it may
be an attractive solution, but identify several potentially serious drawbacks on both observational
and theoretical grounds.  In particular, we argue that previously published estimates of horizontal turbulence
would be sufficient to suppress the instability and that the naturally expected depth of mixing and trends
with luminosity may be in conflict with observational trends.

\section{General Criteria for Convective Instability}

In the presence of a positive $\mu$-gradient $d\mu/dr$ (a negative $\nabla_\mu$), 
a fluid element displaced vertically upwards
will find itself surrounded by material with a higher $\mu$. Whether the fluid element will continue
to rise depends on how efficiently it can exchange mass and heat with its surroundings.
It is easy to anticipate that heat diffusion will favor the instability
by reducing the temperature gradient inside the rising fluid element $\nabla'\equiv (d\ln T'/d\ln P)$,
as compared to its adiabatic value $\nabla_{\rm ad}$, and slightly increasing the gradient in the surrounding 
medium $\nabla\equiv (d\ln T/d\ln P)$, as compared to its radiative value $\nabla_{\rm rad}$ in the absence of mixing. 
In a radiative zone, we always have $\nabla_{\rm rad}\leq \nabla < \nabla' < \nabla_{\rm ad}$. On the other hand,
molecular diffusion and horizontal turbulent diffusion (if the latter is present it will also contribute to
heat diffusion) will decrease the $\mu$-contrast between the fluid element and its
surroundings $|\Delta\mu| = |\mu' (r)-\mu (r)|$, thus hindering the development of convection.
Current theories of rotational mixing predict the existence of strong horizontal turbulence 
(\citealt{chz92,m03,mpz04}); even if the vertical turbulence is
not strong enough to produce mixing such a horizontal turbulence could strongly impact 
an instability driven by differences in composition.
Let us carry out a general investigation of effects produced by these diffusion processes
on the convective instability of the radiative zone in the presence of a $\mu$-gradient inversion.
The magnitude of these effects depends on the P\'{e}clet number that represents
a ratio of thermal and mixing time scale
$$
Pe = \frac{\tau_{\rm th}}{\tau_{\rm mix}}\approx\frac{lv}{(K+D_{\rm h})}\left(\frac{d}{l}\right)^2,
$$
where $\tau_{\rm th}\approx d^2/(K+D_{\rm h})$ and $\tau_{\rm mix} \approx l/v$.
Here, we consider a possibility that the convective motions 
may be organized in elongated narrow structures resembling ``fingers'' whose length $l$ is much larger than their
diameter $d$. In this case, the fluid element velocity $v$, thermal diffusivity $K$, and
horizontal turbulent diffusivity $D_{\rm h}$ should appropriately be averaged over the finger length scale $l$.
For simplicity, we will represent a ``finger'' by a spherical fluid element of the diameter $d$
that travels the path $l\geq d$ before it dissolves. Modeling a ``finger'' with a cylinder would
introduce factors of order unity in our derived relations.
In the limit of a high P\'{e}clet number, mixing is so fast compared to heat exchange that $T'$ undergoes
nearly adiabatic changes, hence $\nabla'\approx\nabla_{\rm ad}$. In this limit, when the rising fluid element
dissolves it has a lower temperature than its surroundings therefore it reduces $T$ locally making $\nabla$ steeper
and closer to $\nabla_{\rm ad}$. In the opposite limit of a low P\'{e}clet number, mixing is so slow that
the radiative and turbulent heat flux from the surroundings can warm up the fluid element thoroughly as it rises.
This brings $\nabla'$ close to $\nabla_{\rm rad}$ while $\nabla$ remains nearly equal to $\nabla_{\rm rad}$
because the fluid element will have $T'\approx T$ when it dissolves.

In our further analysis we will use simple heuristic relations obtained in the mixing length approximation
by \cite{m95} and \cite{tz97} in their investigations of shear instabilities in rotating stars.
We admit that the radiative zone may host both the $^3$He-driven convection and some other extra mixing of a nonconvective origin
(e.g., rotation-driven turbulent diffusion or meridional circulation) at the same time.
Following \cite{z92}, we assume that rotation-driven turbulence
is highly anisotropic, with horizontal components of the turbulent viscosity
strongly dominating over those in the vertical direction, $D_{\rm h}\gg D_{\rm v}$.
We consider a process with a diffusion coefficient $D_{\rm mix} = vl/3$.  At the present stage we leave 
$D_{\rm mix}$ unspecified, and simply solve for the coupled diffusion of heat
and chemicals to evaluate the range of diffusion coefficients over which an instability occurs.  
In the next section we compare with specific (and previously
published) estimates of diffusion coefficients.

The degree to which convection modifies the thermal 
structure of the radiative zone depends on its heat transport efficiency 
\bea
\Gamma = \frac{\Delta E_{\rm trans}}{\Delta E_{\rm ex}} = \frac{\nabla'-\nabla}{\nabla_{\rm ad}-\nabla'} = 
\frac{D_{\rm mix}}{2(K+D_{\rm h})}\left(\frac{d}{l}\right)^2 = \frac{Pe}{6}.
\label{eq:gamma}
\eea
The quantity $\Gamma$  measures the deficit or excess of energy transported by rising or sinking turbulent 
elements $\Delta E_{\rm trans}$ with respect to the energy $\Delta E_{\rm ex}$
the elements gain or lose through the radiative ($K$) 
plus turbulent ($D_{\rm h}$) heat exchange with the surroundings.

For an ideal gas, the square of the Brunt-V\"{a}is\"{a}l\"{a} (buoyancy) frequency is
\bea
N^2 = \frac{g}{H_P}\,(\nabla' - \nabla + \nabla_\mu - \nabla'_\mu).
\label{eq:n2}
\eea
This equation takes into account that horizontal diffusion (molecular plus turbulent) may change
the mean molecular weight of the fluid element $\mu'$ during its motion, so that $\nabla'_\mu\not= 0$.
The convective instability sets in when $N^2$ becomes negative. 
By analogy with equation (\ref{eq:gamma}) and following \cite{tz97}, we introduce a $\mu$-transport efficiency
\bea
\Gamma_\mu = - \frac{\nabla'_\mu - \nabla_\mu}{\nabla'_\mu} = \frac{3D_{\rm mix}}{2(\nu_{\rm mol} +D_{\rm h})}\left(\frac{d}{l}\right)^2,
\label{eq:gammamu}
\eea
where $\nu_{\rm mol}$ is the molecular diffusivity.
Supplementing the radiation luminosity with a contribution to heat transport by convection
in the same manner as \cite{m95} dealt with shear mixing,
we obtain the following relation between $\nabla$, $\nabla_{\rm rad}$ and $\nabla_{\rm ad}$:
\bea
\nabla = \frac{\nabla_{\rm rad} + 6\frac{\Gamma^2}{\Gamma +1}\frac{K+D_{\rm h}}{K}\left(\frac{l}{d}\right)^2\nabla_{\rm ad}}
{1+6\frac{\Gamma^2}{\Gamma +1}\frac{K+D_{\rm h}}{K}\left(\frac{l}{d}\right)^2}.
\label{eq:nablas}
\eea

Combining eqs. (\ref{eq:gamma}\,--\,\ref{eq:nablas}), the instability condition $N^2 < 0$ can be transformed into
\bea
-\nabla_\mu > \left(\frac{\Gamma_\mu + 1}{\Gamma_\mu}\right)
\frac{\Gamma}{1+\Gamma+6\Gamma^2\,\frac{K+D_{\rm h}}{K}\left(\frac{l}{d}\right)^2}\,\,(\nabla_{\rm ad} - \nabla_{\rm rad}).
\label{eq:conv}
\eea

\subsection{The Adiabatic Limit}

In the case of $\Gamma\gg 1$, equations (\ref{eq:gamma}) and (\ref{eq:nablas}) give
$\nabla'\approx\nabla\approx\nabla_{\rm ad}$, which we call ``the adiabatic limit''.
From equations (\ref{eq:gamma}) and (\ref{eq:gammamu}) it follows that 
\bea
\Gamma_\mu = 3\,\frac{K+D_{\rm h}}{\nu_{\rm mol} +D_{\rm h}}\,\Gamma,
\label{eq:relgammas}
\eea
hence $\Gamma_\mu\gg 1$ as soon as $\Gamma\gg 1$ because $\nu_{\rm mol}\ll K$ (Fig.~\ref{fig:f3}).
Using these constraints, the condition (\ref{eq:conv}) is transformed to
\bea
D_{\rm mix} > \frac{1}{3}\,K\,\frac{\nabla_{\rm ad}-\nabla_{\rm rad}}{|\nabla_\mu|}\gg K.
\label{eq:adiab}
\eea
Radial displacements $l\sim 10^8$ cm of fluid elements with velocities of order $5\times 10^4$ cm\,s$^{-1}$
observed by EDL06 in their 3D red giant model above the major H-burning shell correspond to
a diffusion coefficient $D_{\rm mix}\sim\frac{1}{3}vl\sim 2\times 10^{12}$ cm$^2$\,s$^{-1} \gg K$.
The same or even higher order of magnitude estimate can be obtained if we calculate 
$D_{\rm mix}\sim vH_P$, where $v^2\sim gH_P\Delta\mu/\mu$, as proposed by EDL06 (see next section).
So, it appears that the $^3$He-driven mixing in the 3D red giant model
of EDL06 somehow wound up in the metastable adiabatic limit.
We explore the consequences of such a rapid mixing process for both surface abundances and the thermal structure in section 4.

\subsection{The Radiative Limit}

\begin{figure}
\plotone{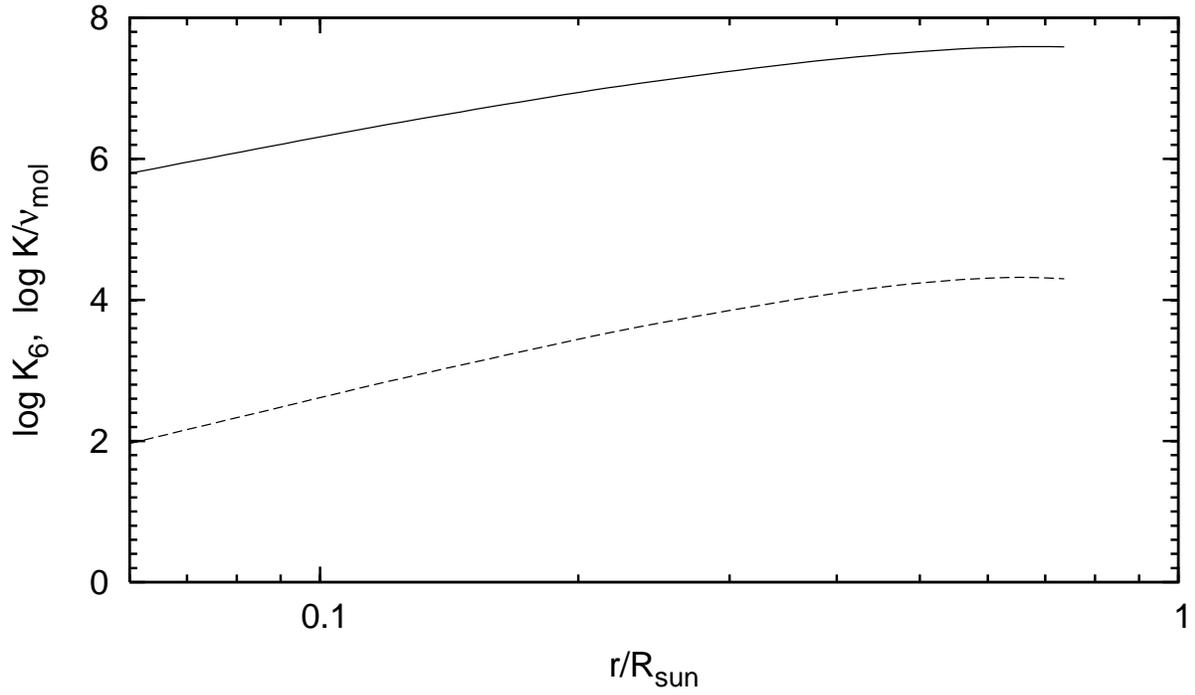}
\caption{Profiles of the ratio of the thermal and molecular diffusivity (solid curve) and of
         the quantity $K_6\equiv K/10^6$ cm$^2$\,s$^{-1}$ (dashed curve) in the radiative zone of
         our $0.83\,M_\odot$ bump luminosity model.
         } 
\label{fig:f3}
\end{figure}

Let us now consider a situation when $\Gamma(l/d)^2\ll 1$. This also means that $\Gamma\ll 1$ because we have assumed that $d\leq l$.
In this case, equations (\ref{eq:gamma}) and (\ref{eq:nablas}) give $\nabla'\approx\nabla\approx\nabla_{\rm rad}$, therefore
we will refer to it as ``the radiative limit''. Given that in equation (\ref{eq:relgammas})
the ratio $K/\nu_{\rm mol}\gg 1$ (solid curve in Fig.~\ref{fig:f3}), the assumption that $\Gamma\ll 1$ does not automatically
lead to $\Gamma_\mu < 1$ unless $D_{\rm h}\ga K$. Hence, we have to consider the cases of $\Gamma_\mu\gg 1$ and
$\Gamma_\mu < 1$ separately.

a). In the radiative limit ($\Gamma\ll 1$), values of $\Gamma_\mu\gg 1$ can be met only if $D_{\rm h}\ll K$. 
Using these constraints, the condition (\ref{eq:conv}) is simplified to
\bea
D_{\rm mix} < \frac{2K}{\nabla_{\rm ad}-\nabla_{\rm rad}}\,|\nabla_\mu|\left(\frac{l}{d}\right)^2,
\label{eq:dmix}
\eea
The right-hand side of (\ref{eq:dmix}) adequately reproduces
both the diffusion coefficient for thermohaline convection derived by \cite{kea80}
\bea
D_{\rm Kipp} = \frac{3K}{\nabla_{\rm ad}-\nabla_{\rm rad}}\,|\nabla_\mu|,
\label{eq:dkipp}
\eea
who argued that $l$ should be of order $d$, and the rate of mixing by elongated narrow ``fingers'' ($l>d$)
advocated by \cite{u72}
\bea
D_{\rm Ulrich} = \frac{8}{3}\pi^2\,\frac{K}{\nabla_{\rm ad}-\nabla_{\rm rad}}\,|\nabla_\mu|\left(\frac{l}{d}\right)^2.
\label{eq:dulrich}
\eea
Thus, in the radiative limit with $D_{\rm h}\ll K$ we can readily identify mixing with thermohaline convection.
Substituting expressions (\ref{eq:dkipp}\,--\,\ref{eq:dulrich}) in place of $D_{\rm mix}$, we find that, for thermohaline convection
driven by the $^3$He burning,
\bea
\Gamma \approx \frac{vl}{6K}\,\left(\frac{d}{l}\right)^2 \approx
\frac{D_{\rm mix}}{2K}\,\left(\frac{d}{l}\right)^2 \approx
\frac{|\nabla_\mu|}{\nabla_{\rm ad}-\nabla_{\rm rad}} \ll 1,
\label{eq:gamma2}
\eea
as we assumed.

b). If $D_{\rm h}$ is not negligibly small compared with $K$ then we sure have $D_{\rm h}\gg\nu_{\rm mol}$, and relation
(\ref{eq:relgammas}) can be re-written as
\bea
\Gamma_\mu = 3\,\frac{K+D_{\rm h}}{D_{\rm h}}\,\Gamma.
\label{eq:relgammas2}
\eea
For $\Gamma_\mu < 1$, the condition (\ref{eq:conv}) takes a form
\bea
\frac{|\nabla_\mu|}{\nabla_{\rm ad}-\nabla_{\rm rad}} > \frac{1}{3}\frac{D_{\rm h}}{K+D_{\rm h}}
\label{eq:gmlt1}
\eea
(compare it with condition 5 from \citealt{v04}).
In the radiative zone of an LM-RGB star, the left-hand side of (\ref{eq:gmlt1}) is of order $10^{-3}$ at most.
A profile of the quantity $K$ in the radiative zone of our $0.83\,M_\odot$ bump luminosity model
is plotted with dashed curve in Fig.~\ref{fig:f3}. Given that $K\la 10^9$ cm$^2$\,s$^{-1}$ for
$r\la 0.1\,R_\odot$ and comparing the ratio $D_{\rm h}/K$ with the number $10^{-3}$, 
we conclude that the horizontal turbulent diffusion
with $D_{\rm h} \ga 10^6$ cm$^2$\,s$^{-1}$ may hinder the development of convective instability.
Interestingly that such values of $D_{\rm h}$ have been found by \cite{pea06} in their low-metallicity
$0.85\,M_\odot$ bump luminosity model even for the less favorable case of solid-body rotation
of the convective envelope. For the case of differential rotation of the convective envelope,
which is suggested by observed fast rotation of horizontal branch stars, they have obtained
$D_{\rm h} \ga 10^7$ cm$^2$\,s$^{-1}$. For such large values of $D_{\rm h}$, the right-hand side of (\ref{eq:gmlt1})
exceeds the expression on the left-hand side at least for $r\la 0.1\,R_\odot$ (Fig.~\ref{fig:f3}), 
therefore the convection (thermohaline mixing) will probably be suppressed there.

\section{Ulrich's Solution in the Presence of Horizontal Turbulent Diffusion}
\label{sec:ulrich}

In this section we demonstrate that \cite{u72} could have come to a conclusion about the suppression
of thermohaline convection by the strong horizontal turbulent diffusion similar to that made by us
(condition \ref{eq:gmlt1}) if he had included $D_{\rm h}$ in his equations. To do this, we start with
the linearized equations of conservation of momentum, thermal energy, and chemical composition
similar to those used by Ulrich (his equations 1\,--\,3) but with the diffusion coefficient
$D_{\rm h}$ taken into account
\bea
\label{eq:ulrich1}
2\,\frac{dv}{dt} & = & -\frac{\nu}{d^2}\,v - g\,\delta\ln T + g\,\delta\ln\mu, \\
\label{eq:ulrich2}
\frac{d\,\delta\ln T}{dt} & = & \frac{(\nabla_{\rm ad}-\nabla_{\rm rad})}{H_P}\,v - \frac{(K+D_{\rm h})}{d^2}\,\delta\ln T, \\
\label{eq:ulrich3}
\frac{d\,\delta\ln\mu}{dt} & = & -\frac{\nabla_\mu}{H_P}\,v - \frac{(\nu_{\rm mol}+D_{\rm h})}{d^2}\,\delta\ln\mu. \\ \nonumber
\eea
Here, $\nu = \nu_{\rm mol} + \nu_{\rm rad}$ is the total (molecular plus radiative) viscosity,
$\delta\ln T = \ln T'(r) - \ln T(r)$, $\delta\ln\mu = \ln\mu'(r) - \ln\mu (r)$, other symbols having been
defined previously.

The characteristic polynomial for the linear system of ODEs (\ref{eq:ulrich1}\,--\,\ref{eq:ulrich3})
can be written in the following form:
\bea
\label{eq:dispersion}
\left(\omega\tau_{\rm th}\right)^3 + \left[1+\left(\tau_{\rm th}/\tau_\nu\right)+\left(\tau_{\rm th}/\tau_\mu\right)\right]\,\left(\omega\tau_{\rm th}\right)^2 + & \\ \nonumber
\left[\left(\tau_{\rm th}/\tau_\nu\right)+\left(\tau_{\rm th}/\tau_\nu\right)\left(\tau_{\rm th}/\tau_\mu\right)+
\left(\tau_{\rm th}/\tau_\mu\right)+\left(\tau_{\rm th}^2N^2\right)\right]\,\left(\omega\tau_{\rm th}\right) + & \\ \nonumber
\left[\left(\tau_{\rm th}/\tau_\nu\right)\left(\tau_{\rm th}/\tau_\mu\right)+\left(\tau_{\rm th}^2N_\mu^2\right)+ 
\left(\tau_{\rm th}/\tau_\mu\right)\left(\tau_{\rm th}^2N_T^2\right)\right] = 0, & \\ \nonumber
\eea
where $\omega = 2\pi\tau_{\rm mix}^{-1}$ is the eigen frequency of stable ($\omega < 0$) or unstable ($\omega > 0$) mode,
$N_T^2 = g(\nabla_{\rm ad}-\nabla_{\rm rad})\,H_P^{-1}$ and $N_\mu^2 = g\nabla_\mu\,H_P^{-1}$ are the squares of the $T$- and $\mu$-component of
the Brunt-V\"{a}is\"{a}l\"{a} frequency for the ideal gas, $N^2 = N_T^2 + N_\mu^2$, while
$\tau_\nu = d^2/\nu$, $\tau_{\rm th} = d^2/(K+D_{\rm h})$, and $\tau_\mu = d^2/(\nu_{\rm mol}+D_{\rm h})$
denote the viscous, thermal, and horizontal diffusion timescales for the fluid element.
Our equation (\ref{eq:dispersion}) is equivalent to Ulrich's dispersion relation (10).

\begin{figure}
\plotone{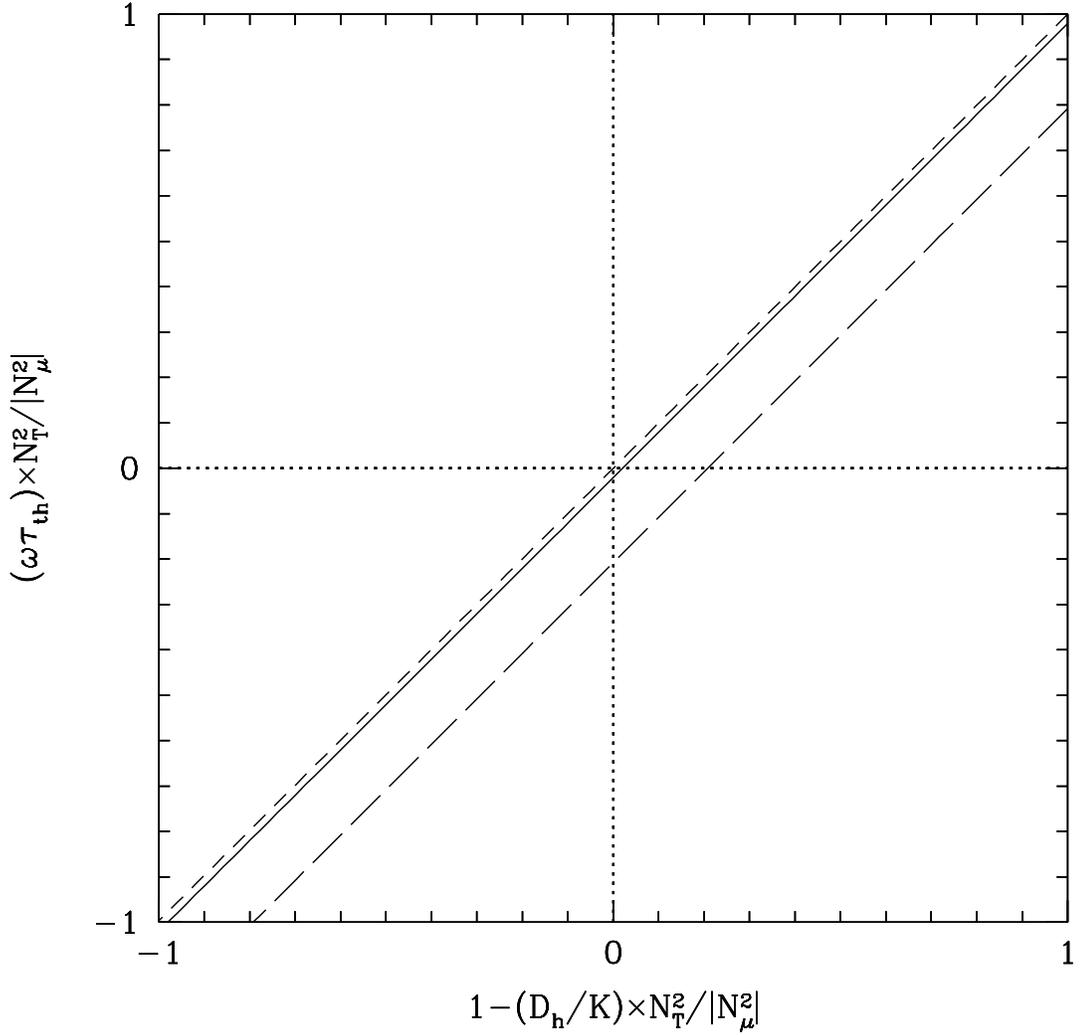}
\caption{Solutions of the dispersion relation (\ref{eq:dispersion}) as a function of
         the horizontal turbulent diffusion coefficient. Both the only real root $\left(\omega\tau_{\rm th}\right)$
         and the parameter $(D_{\rm h}/K)$ have been scaled appropriately to reveal the dependence 
         (\ref{eq:dulrich}) and the instability condition (\ref{eq:gmlt1}). Plotted are
         the solutions for the fluid element diameter $d = 0.01\,H_P$ and three values of
         $N_\mu^2 = -10^{-7}$ (this is a characteristic value for $^3$He burning in low-metallicity bump luminosity
         stars with $M\approx 0.8\,\msol$\,; solid line), $-10^{-6}$ (short-dashed line),
         and $-10^{-8}$ (long-dashed line).
         } 
\label{fig:f6}
\end{figure}

The only real root of the polynomial (\ref{eq:dispersion}) is plotted as a function of $(D_{\rm h}/K)$ in Fig.~\ref{fig:f6} for
the ratio $d/H_P = 0.01$ and three different values of $N_\mu^2 = -10^{-7}$ (solid line), $-10^{-6}$ (short-dashed line),
and $-10^{-8}$ (long-dashed line). The first value of $N_\mu^2$ is close to the minimum one found in the region of the
$\mu$-gradient inversion produced by $^3$He burning in a low-metallicity bump luminosity star with a mass $M\approx 0.8\,\msol$. The quantities $\left(\omega\tau_{\rm th}\right)$
and $(D_{\rm h}/K)$ in Fig.~\ref{fig:f6} have been scaled appropriately to reveal both our guessed functional dependence (\ref{eq:dulrich}) and
instability condition (\ref{eq:gmlt1}). At a fixed value of $N_\mu^2$ our solutions for the ratio $d/H_P$ varying from
$0.01$ down to $0.0001$ coincide. From Fig.~\ref{fig:f6}, we conjecture that
\bea
D_{\rm mix} \sim \frac{l^2}{\tau_{\rm mix}} \sim \frac{l^2}{\tau_{\rm th}}\,\left(\omega\tau_{\rm th}\right) \sim K\,\frac{|N_\mu^2|}{N_T^2}\,\left(\frac{l}{d}\right)^2\times
\left(1 - \frac{D_{\rm h}}{K}\,\frac{N_T^2}{|N_\mu^2|}\right),
\label{eq:modulrich}
\eea
i.e. the thermohaline convective instability may develop ($\omega > 0$) only if $D_{\rm h} < K|N_\mu^2|/N_T^2$. The latter condition is equivalent
(ignoring factors of order unity) to (\ref{eq:gmlt1}) for $D_{\rm h} < K$. If $D_{\rm h}\ll K$ then we can neglect the term in the parentheses.
In this case we obtain an expression for $D_{\rm mix}$ similar to Ulrich's original formula.

Note that an equation similar to (\ref{eq:modulrich}) can be derived directly from the dispersion relation
(\ref{eq:dispersion}) in the same way that Ulrich used to estimate $D_{\rm mix}$ in (\ref{eq:dulrich}).
Following him, we ignore the viscosity and take advantage of the fact that the thermohaline mode is 
the slowest one ($\omega\tau_{\rm th}\sim\tau_{\rm th}/\tau_{\rm mix}\ll 1$).
Therefore, we can neglect the quadratic and cubic terms in (\ref{eq:dispersion}) as well as terms with $\tau_\nu^{-1}$. 
Taking into account that $N_T^2\gg |N_\mu^2|$, and $N^2 \gg \left(\tau_{\rm th}\tau_\mu\right)^{-1}$, we finally obtain
\bea
D_{\rm mix} \sim (K+D_{\rm h})\,\frac{|N_\mu^2|}{N_T^2}\,\left(\frac{l}{d}\right)^2\times
\left[1 - \frac{(\nu_{\rm mol}+D_{\rm h})}{(K+D_{\rm h})}\,\frac{N_T^2}{|N_\mu^2|}\right],
\label{eq:modulrich2}
\eea
which is reduced to (\ref{eq:modulrich}) if $\nu_{\rm mol} < D_{\rm h} < K$.

\section{Observational Constraints on the Extra-Mixing Rate}
\label{sec:mixing}

The $^3$He mechanism has a different underlying origin than rotational mixing,
and it is therefore useful to re-evaluate the global implications of its operation.
This is especially true because we have identified two stable branches with very
different timescales.  We begin by establishing that a mixing process which occurs over
too short of a timescale would have a dramatic impact on the thermal structure which contradicts
the observational data.  We also demonstrate that diffusion coefficients derived from empirical
mixing estimates are consistent with the data.

We then critically examine whether thermohaline
mixing is capable of reproducing the data, and the answer depends critically on the assumed geometry
of the fluid elements (and the rate at which they can achieve thermal balance with their surroundings).
However, we can also extend the same underlying mechanism to predict luminosity trends, behavior in
other evolutionary states, and implications for interacting binaries.  Definite predictions emerge, and
we can identify both existing conflicts and topics which will require further calculation.
In particular, we contend that the natural expectation would be for mixing that is shallower
in temperature and weaker for bright giants than current data indicates.  The predictions for other types
of stars are fundamentally different, and in our view the latter test will ultimately prove to be decisive.

\subsection{Rapid Mixing in the Adiabatic Limit}

The observed changes in the surface abundances of Li and C and in the \cc\ ratio 
as a function of luminosity in LM-RGB stars (e.g., \citealt{chea98,grea00}) can be used to constrain the depth and rate of
extra mixing in them. If $\Delta\log T$ is the difference between the logarithms of temperature at the base of the
H-burning shell and at the maximum depth of extra mixing and $D_{\rm mix}$ is the diffusion coefficient then,
as \cite{dv03} have demonstrated, extra mixing in LM-RGB stars can be
parameterized by any pair of correlated values within the close limits
specified by $\Delta\log T\approx 0.19$ and $D_{\rm mix}\approx 4\times 10^8$\,cm$^2$\,s$^{-1}$, to
$\Delta\log T\approx 0.22$ and $D_{\rm mix}\approx 8\times 10^8$\,cm$^2$\,s$^{-1}$.  
However, that parameterization
did not take into consideration the thermal response of the radiative zone
to mixing. 

In this paper, we use a more consistent
parametric prescription by letting $D_{\rm mix}$ be equal to a fixed fraction of $K$
and allowing the temperature gradient in the radiative zone to be modified as prescribed by 
the mixing length theory relation (\ref{eq:nablas}).
It is inspired by a similarity between
equation (\ref{eq:dkipp}) and the functional dependence of $D_{\rm mix}$ on $K$ obtained for 
rotational shear mixing by \cite{mm96}.

\begin{figure}
\plotone{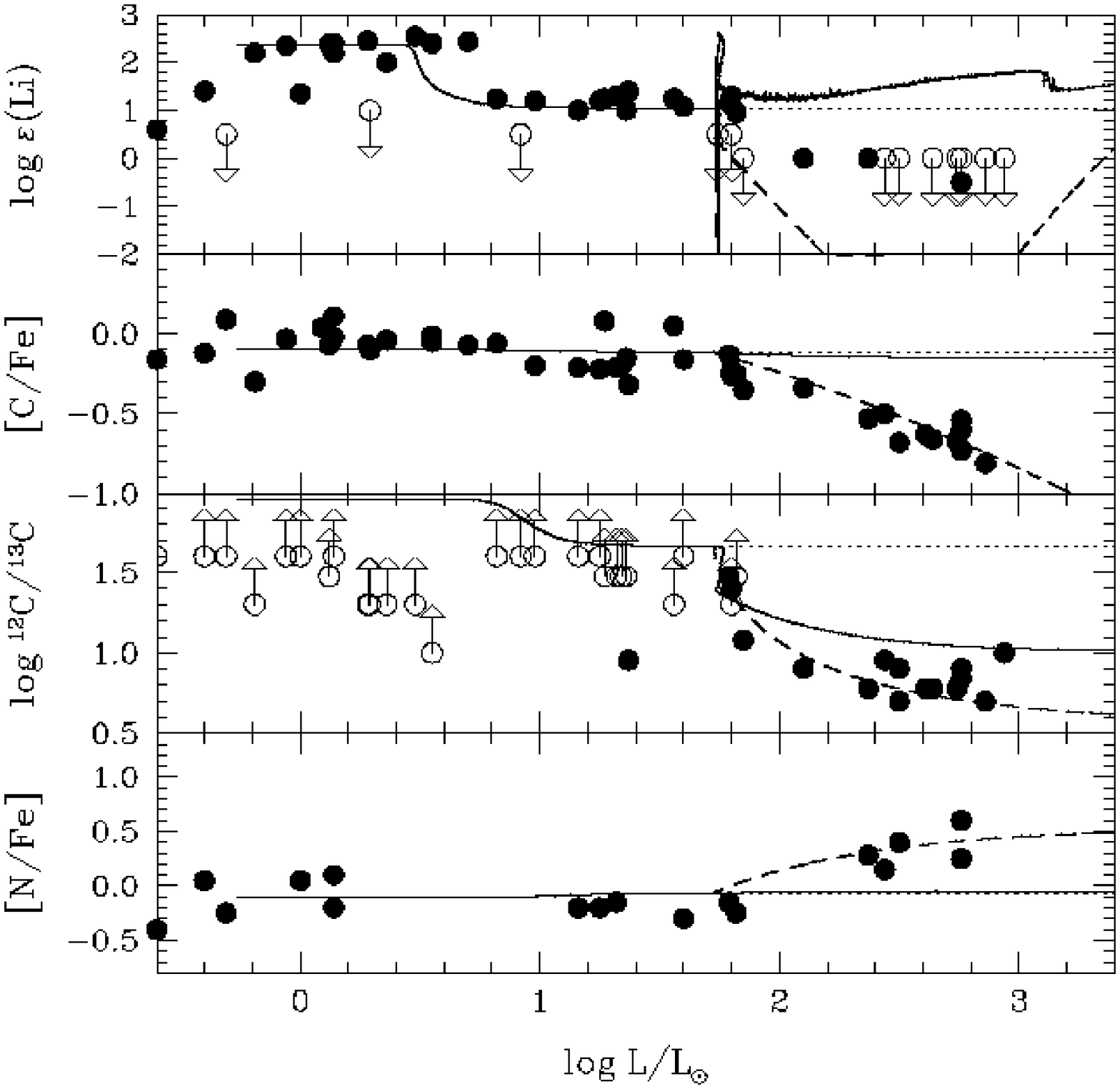}
\caption{Comparison of the observational data from \cite{grea00}
         for field metal-poor ($-2\la\mbox{[Fe/H]}\la -1$) low-mass stars
         (circles) with results of our computations of the evolution of the $0.83\,\msol$ star
         with the $^3$He-driven mixing (solid curves, equations \ref{eq:dmu})
         and extra mixing with the rate $D_{\rm mix} = 0.02\,K$ and depth
         $\Delta\log T = 0.19$ (dashed curves). Dotted lines show 
         predictions of the standard theory. For further details, see text.
         } 
\label{fig:f2}
\end{figure}

We derive $\Gamma = 0.01$ for the mixing depth $\Delta\log T=0.19$ constrained by \cite{dv03}, which results in
the diffusion coefficient $D_{\rm mix}=0.02\,K$ (assuming that $l\approx d$ and $D_{\rm h}\ll K$ in equation \ref{eq:gamma}).
These models reproduce quite well the chemistry of LM-RGB stars
above the bump luminosity (dashed curves in Fig.~\ref{fig:f2}). They do not         
noticeably change our models' photometric behavior and evolutionary time scale near the bump luminosity
compared to models without extra mixing. This is important because
photometric observations of LM-RGB stars, in particular the absolute V-band magnitude of the luminosity bump
and the excess number of stars in the bump, agree with predictions of standard stellar evolution
theory (e.g., \citealt{bch06}). Our test computations have shown that the ratio of the gradients
in equations (\ref{eq:dkipp}\,--\,\ref{eq:dulrich}) does not change much with radius
and that it is roughly proportional to the abundance of $^3$He left in the mixing zone. 

On the other hand, already at $\Gamma = 0.4$ the bump luminosity
zigzag gets so extended toward a lower luminosity (dashed curve in upper panel in Fig.~\ref{fig:f1}), and the model star spends
so long time following it that this peculiar behavior would sure have been noticed in
photometric studies, such as the counting of the number densities of stars as a function of luminosity on the RGBs of
globular clusters. To be more specific, it takes about twice as long for the model star
to make the extended zigzag as compared to the standard evolution. It should also be noted
that the model of such a rapidly mixed star spends most of this time residing near  the bottom of the zigzag, about
0.3 magnitude below the standard bump luminosity. Values of $\Gamma > 0.4$ would result in
even more drastic changes. This behavior is similar to that
of models of rapidly rotating RS CVn binaries found by \cite{denea06} except that
in the latter case the extended zigzag was produced by an increase of $\nabla$ caused
by the stars' rotational deformations.

The estimate of the turbulent velocity $v^2\sim gH_P(\Delta\mu/\mu)$ for the $^3$He-driven mixing
used by EDL06 can be obtained from equation (\ref{eq:n2}) if we put into it $\nabla' - \nabla = \nabla'_\mu = 0$.
Indeed, in this case we can approximately consider that
$v^2\sim H_P^2\,|N^2|=gH_P|\nabla_\mu|=gH_P^2(d\ln\mu/dr)\sim gH_P(\Delta\mu/\mu)$, at least in the vicinity of
the $^3$He-burning shell where the mean molecular weight height scale is of order $H_P$ (bottom panel in Fig.~\ref{fig:f1}).
Note that in the mixing length theory the approximations $\nabla' = \nabla$ 
and $\nabla'_\mu = 0$ are correct only in the limits of $\Gamma = \infty$
or $\Gamma = 0$, and $\Gamma_\mu = \infty$ (eqs. \ref{eq:gamma}\,--\,\ref{eq:nablas}). 

Neglecting the influence of extra mixing on the radiative zone's thermal stratification,
we have computed the evolution of our model also with the following diffusion coefficient:
\bea
D_{\rm mix}=\frac{1}{3}H_Pv,\ \ \mbox{where}\ \ v^2=\frac{1}{8}gH_P|\nabla_\mu|.
\label{eq:dmu}
\eea
The factor $\frac{1}{8}$ comes from the mixing length theory (\citealt{wea04}, Ch.~14).
The depth of this ``$^3$He-driven'' extra mixing has been determined by locating the minimum of $\mu$
above the H-burning shell. Outside of this point, $\mu$ increases with radius due to the $^3$He burning and
mixing. We think that our prescription is in line with that EDL06 had in mind.
It should be noted that in our computations values of $D_{\rm mix}$ were determined at each time step
using a current distribution of $\mu$ that was constantly modified by extra mixing.
Characteristic values of $D_{\rm mix}$ obtained in this self-regulating way were of order $10^{12}$\,cm$^2$\,s$^{-1}$.
Such fast extra mixing is known to produce large amounts of $^7$Li via the Cameron-Fowler
mechanism (e.g., \citealt{dw00,dh04}). This disagrees with the low (often undetectable) Li abundances in the majority of
LM-RGB stars located above the bump luminosity (compare the solid curve above $\log\,L/L_\odot\approx 1.8$ 
with the observational data points in top panel in Fig.~\ref{fig:f2}).
In accordance with EDL06, we did find a modest decrease in the \cc\ ratio.
However, it is obvious that the observed evolutionary changes of the surface chemical composition
of LM-RGB stars require a slightly deeper (in order to reproduce the C depletion)
and much slower (in order to keep the Li abundance low) extra mixing than that advocated by EDL06.
Besides, extra mixing with diffusion coefficients $D_{\rm mix}\gg K\sim 10^8$\,--\,$10^{10}$\,cm$^2$\,s$^{-1}$
would bring the radiative zone to the quasi-adiabatic state (unless the fluid elements
have finger-like structures, which has not been reported by EDL06), which would cause the star to make
a prolonged excursion below the bump luminosity in contradiction with observations.
Therefore, we believe that the rapid mode originally invoked by EDL06 does not operate,
but that mild mixing is indicated.

\subsection{Mild Thermohaline Mixing in the Radiative Limit}

Because the $^3$He-driven thermohaline convection is expected to work in the radiative limit it is
interesting to test if it is fast and deep enough to
explain extra mixing in LM-RGB stars. In Fig.~\ref{fig:f4}, we illustrate characteristic diffusion
coefficients. 
In order to produce mixing, a physical mechanism must operate over a timescale shorter than
the inflow rate.
The dot-dashed curve shows our empirically constrained diffusion coefficient $D_{\rm mix}=0.02\,K$. 
For comparison, the dashed curve shows a minimum threshold diffusion coefficient
$D_{\rm inflow}=|\dot{r}|H_P$, where $|\dot{r}|$ is a mass inflow rate of H-rich material
that flows from the bottom of convective envelope toward the H-burning shell (resembling a spherical
accretion).
The bottom solid curve shows a profile of
the diffusion coefficient (\ref{eq:dkipp}) in our unmixed bump luminosity model.
Once mixing ensues, the $\mu$-gradient inversion spreads out over the entire radiative zone
above the $^3$He-burning shell.
The final state is illustrated with the top solid line.
We conclude that, in the prescription given by \cite{kea80}, 
the thermohaline convection could marginally commence and mix a narrow region in the vicinity of the $^3$He-burning
shell (at $r\approx 0.08\,R_\odot$ in Fig.~\ref{fig:f4}). However, the Kippenhahn et al.
diffusion coefficient is only marginally large enough to trigger this process and is two orders of
magnitude below the empirical value.

If we adopt the Ulrich prescription (\ref{eq:dulrich}) with
$l\approx 10\,d$, our diffusion coefficient is raised by a factor of $10^2$ and the mechanism
may be viable. This is the approach advocated by \cite{chz07} (incidentally, their paper
was posted on astro-ph on the same day when we submitted the first version of our paper).

\begin{figure}
\plotone{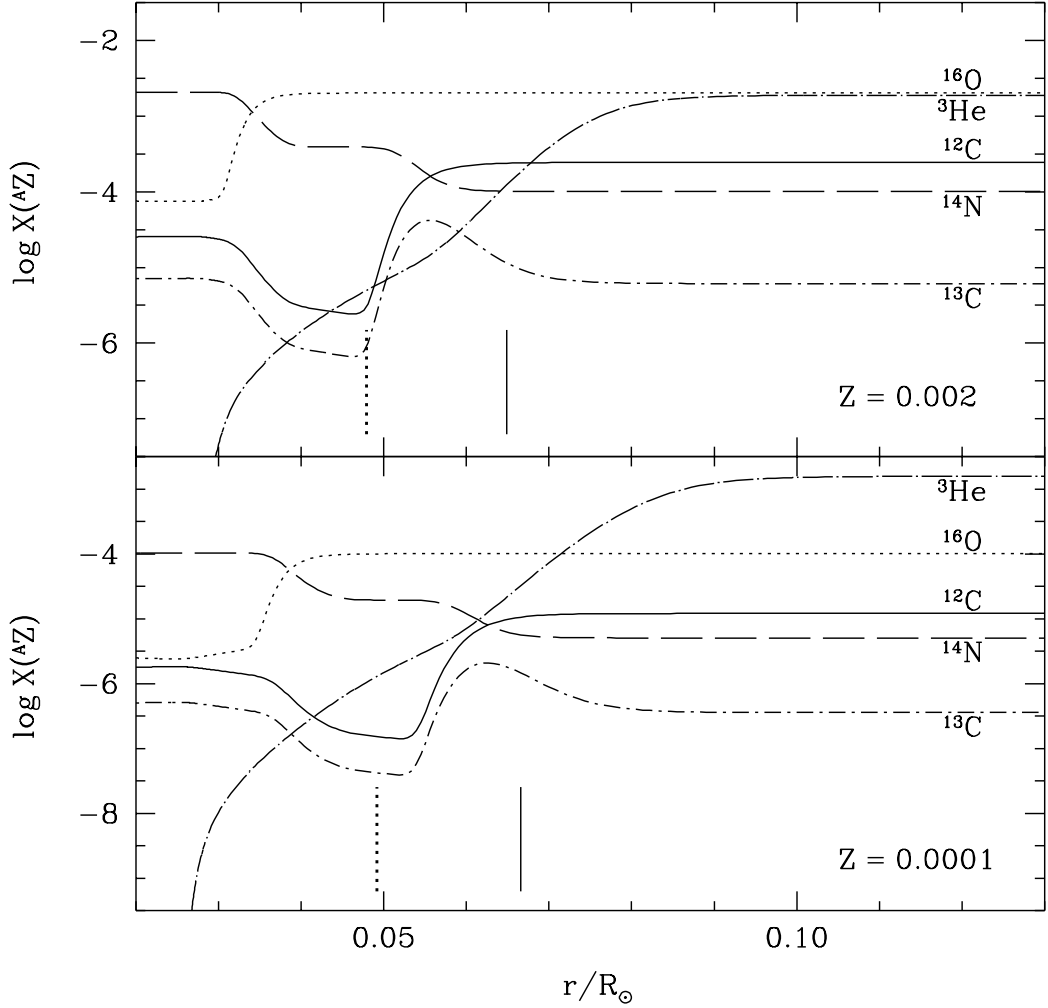}
\caption{Profiles of the mass fractions of $^3$He and CNO elements in the vicinity of
         the H-burning shell in $0.83\,M_\odot$ bump luminosity models with
         the heavy-element mass fractions $Z=0.002$, and $Z=0.0001$. Vertical solid
         line segments show locations of the minimum on the $\mu$-profile. Dotted line
         segments are placed one pressure scale height below $\mu_{\rm min}$.
         Without overshooting, the depth of the $^3$He-driven thermohaline convection
         would be at the locations of the solid segments. There would be no evolutionary
         C depletion in this case, contrary to observations.
         } 
\label{fig:f5}
\end{figure}

Unfortunately, \cite{chz07} do not explain how they have chosen the depth of mixing.
We find that the depth corresponding to a minimum on the $\mu$-profile (solid vertical line segments in Fig.~\ref{fig:f5})
is not sufficient to produce the observed C depletion (top second panel in Fig.~\ref{fig:f2}).
An overshooting on a length scale of order $H_P$ could do it (dotted vertical line
segments in Fig.~\ref{fig:f5} are placed at a distance $H_P$ below $\mu_{\rm min}$) but then
thermohaline ``fingers'' would have to penetrate down a region of higher $\mu$ where they
should experience a strong breaking. It should also be noted that the penetration of
a region with the negative $d\mu/dr$ below $\mu_{\rm min}$ would reduce the average mixing rate by
decreasing the slope of the positive $d\mu/dr$ in the mixing zone. Given these
uncertainties that cannot be resolved from first principles but instead require empirical
calibrations and/or higher resolution 3D hydrodynamic simulations, 
we postpone the use of equation (\ref{eq:dulrich}) to our future paper.
 
\section{Concluding Remarks}
\label{sec:concl}

In this paper, we have shown that the $^3$He burning in the radiative zone of an LM-RGB star
may drive convective fluid motions
provided that their heat transport efficiency is either very high (the adiabatic limit) or
extremely low (the radiative limit).
Confirming the conclusions made by \cite{chz07},
we identify the mixing in the radiative limit with thermohaline convection and we
note that this convection would have a sufficiently high rate to explain the observed mixing pattern in LM-RGB stars
only if fluid elements could travel over length scales exceeding their diameters by a factor of 10 or more.
However, we also find that thermohaline convection may be suppressed by
horizontal turbulence if its associated diffusivity $D_{\rm h}\ga 3K|\nabla_\mu|/(\nabla_{\rm ad}-\nabla_{\rm rad})$.
Such values of $D_{\rm h}$ for rotation-induced horizontal turbulence have been obtained
by \cite{pea06} who used a prescription for estimating $D_{\rm h}$ proposed by \cite{mpz04}. 
Although this prescription may be considered
as a very primitive approximation to a complex physical phenomenon,
a similar heuristic approach has been used
to successfully model mixing and angular momentum transport in radiative zones of massive MS stars (e.g., \citealt{tz97,tea97,m03}).
Of course, we recognize that rigorous 3D hydrodynamic simulations have yet to be done
to support or refute these heuristic models.
In addition to the theoretical issues above, there are significant
empirical challenges for an explanation that relies solely on thermohaline convection.
The same physics should consistently be applied to other phases of evolution or
situations where $\mu$ inversions occur.

If extra mixing in RGB stars is really driven by $^3$He burning then it should
die out by the end of the RGB evolution because of the $^3$He exhaustion.
In this case, the $^3$He-driven extra mixing could not resume working
in the same stars on the asymptotic giant branch (AGB). So, we would expect
the absence of observational signatures of extra mixing in low-mass ($M\la 2\,M_\odot$)
AGB stars unless the mixing in them is of a different nature. 
However, given the similarities in their depth and in the structure of radiative zone
where they operate, it is unlikely that the RGB and AGB mixing
have different physical mechanisms.
Contrary to this prediction, there are observational data
indicating the presence or necessity of operation of extra mixing in these stars
(e.g., \citealt{nbw03,mea06}). 
Moreover, to comply with observations,
the AGB mixing has to penetrate close enough to the H-burning shell
to dredge up material processed in the CN-cycle, like in RGB stars, mimicking the convective
hot-bottom burning that occurs in more massive AGB stars.

\begin{figure}
\plotone{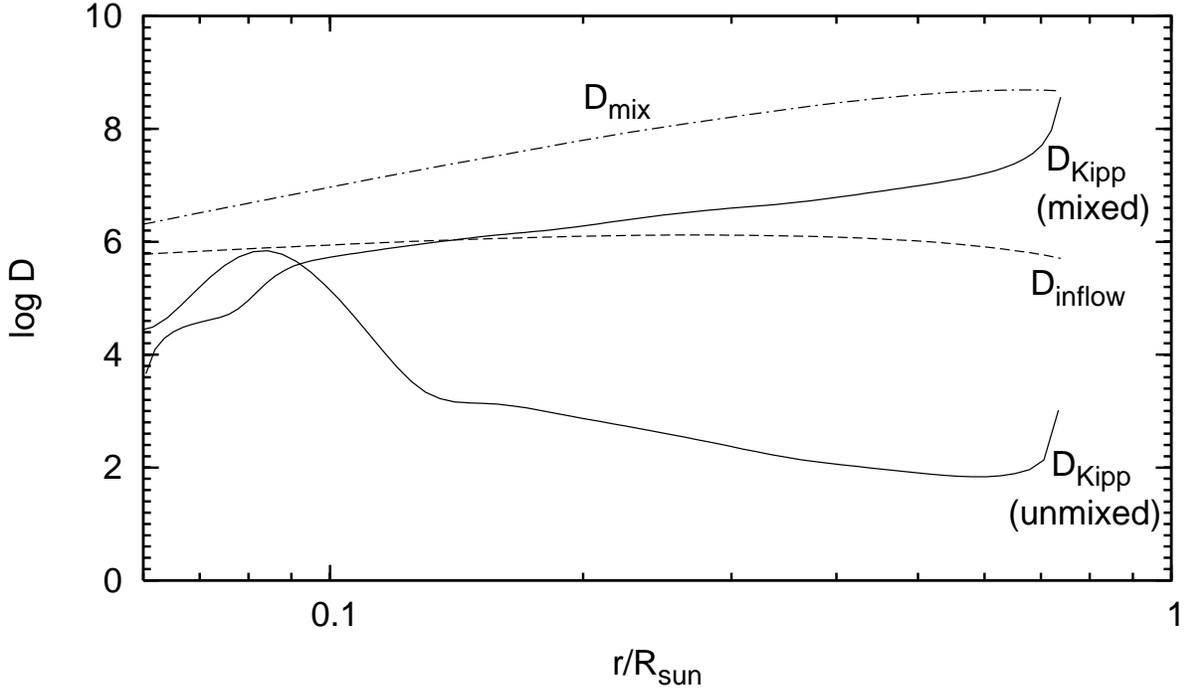}
\caption{Dot-dashed curve -- the empirical profile of $D_{\rm mix} = 0.02\,K$;
         dashed curve -- a minimum threshold profile $D_{\rm inflow}=|\dot{r}|H_P$ that any $D_{\rm mix}$
         must exceed; 
         bottom solid curve -- the $D_{\rm Kipp}$ (thermohaline convection, equation \ref{eq:dkipp}) 
         profile in our unmixed bump luminosity model
         (a hump at $r\approx 0.08\,R_{\rm sun}$ is produced by a local increase of $|\nabla_\mu|$
         in the $^3$He-burning shell);
         top solid curve -- $D_{\rm Kipp}$ in a model in which mixing with $D_{\rm mix}=0.02\,K$
         has spread out the $\mu$-gradient inversion over the entire radiative zone.
         } 
\label{fig:f4}
\end{figure}

\cite{sea07} have used the Kippenhahn et al. prescription (\ref{eq:dkipp}) to model
thermohaline mixing in a metal-poor low-mass
MS star accreting wind material from its AGB binary companion enriched in He and C.
Such accretion is believed to be the primary process responsible for the formation of
the so-called carbon-enhanced metal-poor (CEMP) stars. They have found that thermohaline
convection mixes almost 90\% of the star within about $10^9$ years after the accretion.
On the other hand, the RGB mixing pattern can be reproduced only if the diffusion coefficient given by
equation (\ref{eq:dkipp}) is increased by a factor of $10^2$ to $10^3$ (\citealt{chz07}; also see our Fig.~\ref{fig:f4}).
In this case, thermohaline mixing in CEMP MS stars would dilute the accreted material
on a much shorter time scale of order $10^6$\,--\,$10^7$ years. Unless
most of the CEMP stars accreted substantial fractions of their initial masses,
their rather high frequency $f_{\rm CEMP}\ga$\,20\% among very metal-poor stars (e.g., \citealt{lea06})
would look surprising. Furthermore, both \cite{lea06} and \cite{aea07} have found
anti-correlations between [C/H] (and [(C+N)/H]) and luminosity for Ba-enhanced CEMP stras
spanning over three orders of magnitude in $L$ that they interpreted as 
an evidence of dilution of the envelope material in the accreting companion.
Thermohaline convection on a time scale of order $10^6$\,--\,$10^7$ years
would mix the CEMP MS stars almost instantaneously and well before
their luminosity would begin to increase due to the core H exhaustion.
In that case, the mentioned anti-correlations could not have appeared.
Also note that even on the lower RGB Ba-enhanced CEMP stars have quite low carbon isotopic ratios
($^{12}$C/$^{13}$C\,$ < 20$; \citealt{rea05}) in a striking contrast with the values of
$^{12}$C/$^{13}$C\,$ \gg 1000$ that the low-mass AGB stars are predicted to 
return to the interstellar medium (\citealt{h04}). Extra mixing (in the low-mas AGB stars)
could easily resolve this discrepancy.

The problem of mixing in CEMP MS and RGB stars has recently been addressed by \cite{dp07}.
Particularly, they have shown that the first dredge-up dilution of CN enrichment in CEMP stars relative to 
their MS precursors is indeed a plausible explanation of the observed anticorrelation of [N/Fe]
with $\log\,L/L_\odot$ and that it contradicts models that rely on efficient thermohaline mixing induced by small $\mu$
gradients in red giants. This result has independently been confirmed by \cite{aea08}.
The suppression of thermohaline convection by rotationally driven horizontal turbulence
may explain its reduced efficiency in MS CEMP stars.

Another potentially serious problem for the $^3$He-driven thermohaline convection
could be to explain available observational evidence of enhanced extra mixing in LM-RGB stars.
Firstly, observations show that
in some globular clusters the anti-correlated abundance variations of C and N in red giants
become larger when stars approach the RGB tip. Moreover, extremely large values of
the N abundance in some of these stars indicate the dredge-up of material in which not only C
but also a fraction of O has been converted into N (\citealt{sbh05}).
Secondly, at least in the globular cluster M13, the relative number of upper RGB stars with
the O--Na anti-correlation increases with luminosity (\citealt{jea05}).
Thirdly, \cite{dpt06} have shown that the $^{19}$F abundance variations found in bright
red giants of the globular cluster M4 by \cite{sea05} may also require that extra mixing in them to operate
much faster and somewhat deeper than in LM-RGB stars immediately above the bump luminosity.
It is difficult to interpret these data by the $^3$He-driven mixing because its efficiency should
decline toward the RGB tip in proportion as $^3$He gets depleted.
Unfortunately, the question of evolutionary Na, O, and $^{19}$F abundance variations in
globular-clusters RGB stars is still a matter of debate from the observational point of view.
Therefore, we consider them as a potential rather than a real problem
for the $^3$He thermohaline convection.

A similar problem
is encountered when one tries to understand the phenomenon of Li-rich giants.
There are convincing arguments that high Li abundances in these LM-RGB stars are produced via the Cameron-Fowler
mechanism that also requires enhanced extra mixing with $D_{\rm mix}\sim 10^{10}$\,--\,$10^{11}$ cm$^2$\,s$^{-1}$
(\citealt{dw00,dh04}). It should be noted that most of the Li-rich giants are located above the bump luminosity
(\citealt{chb00}). Besides, their proportion among rapid rotators ($v\sin\,i\geq 8$\,km\,s$^{-1}$)
is $\sim$\,50\% which is considerably larger than $\sim$\,2\% of Li-rich stars among the much more common slowly
rotating ($v\sin\,i\la 1$\,km\,s$^{-1}$) K-giants (\citealt{dea02}). It is not clear why
the $^3$He-driven mixing would be enhanced in fast rotators. Oppositely, \cite{c99} argues
that a larger shear due to differential rotation should decrease the efficiency of mixing by thermohaline convection.
A larger shear would also intensify horizontal turbulence, thus hindering thermohaline convection even stronger.
So, the Li-rich giants seem to support the hypothesis of rotation-induced mixing rather than that of
thermohaline mixing. This may not necessarily be rotational shear mixing that has already been criticized.
Instead, rotation may deposit its kinetic energy to mixing less directly, e.g. through generation of
buoyant magnetic flux tubes (\citealt{bea07}).

For the adiabatic limit, the predicted evolutionary changes of
the surface composition of LM-RGB stars disagree with observational data.
Besides, high $\Gamma$'s would bring the radiative zone to the quasi-adiabatic
state which would result in a photometric behavior of the RGB star inconsistent with the observed one.

In the vicinity of $\mu_{\rm min}$, our empirically constrained diffusion coefficient
has values of order  $D_{\rm mix}\sim 10^6$\,--\,$10^7$ cm$^2$\,s$^{-1}$  (dot-dashed curve in Fig.~\ref{fig:f4}).
If we assume that in real LM-RGB stars extra mixing is produced by thermohaline convection whose
fluid elements have a ratio of $l/d\sim 10$, where $l$ does not exceed
the local pressure scale height $H_P\sim 0.02\,R_\odot$, then we can estimate the elements' characteristic velocities
$v\approx 3D_{\rm mix}/l \sim 2\times 10^{-3}$\,--\,$2\times 10^{-2}\,(H_P/l)$ cm\,s$^{-1}$. EDL06 have found velocities of order
$5\times 10^4$ cm\,s$^{-1}$ in their 3D red giant model. Those would correspond to fluid elements with
diameters from 6 to 60 cm! Interestingly, such small fluid elements would actually be optically
thin because the photon mean free path is $\sim$\,1 cm in this environment. However, we do not believe
that EDL06 could resolve such small finger-like structures. As we mentioned, they have reported fluid element displacements
of order $l\sim 10^8$ cm. Given the discussed inconsistencies in modeling extra mixing in LM-RGB stars
with the $^3$He-driven convection, we conclude that a different mechanism is worth
searching for. It is also obvious that higher resolution 3D hydrodynamic simulations of the $^3$He-driven
mixing are needed to understand what EDL06 have actually witnessed. In particular,          
we qoute here an issue raised by the referee, who seems to be an expert in the field.
``The EDL06 calculations had no mechanism to simulate the turbulent cascade on scales smaller than the zoning
(a subgrid scale model, such as an eddy viscosity treatment), and yet the calculations did not numerically
blow up. To me this says that the finite difference expressions of the {\tt Djehuty} code are themselves
quite diffusive. Thus, I am surprised that this incredibly small mean molecular weight inversion could generate
significant motion without being squelched by the numerical diffusion. My conclusion is that there are some
serious issues that must be addressed about the numerical behavior of the 3D calculations.''

\acknowledgements
We thank the anonymous referee for useful comments and suggestions
that helped us to improve the manuscript.
We acknowledge support from the NASA grant NNG05 GG20G.


\clearpage
\pagestyle{empty}
\voffset=-3 cm

\clearpage

\clearpage

\clearpage
\pagestyle{plaintop}
\voffset=0 cm

\clearpage

\clearpage


\end{document}